\documentclass{article}
\usepackage{spconf,amsmath,amsfonts,graphicx,hyperref,multirow}
\usepackage{booktabs}

\newcommand{\sname}{\emph{\textit{ContrastASC}}}
\newcommand{\sssec}[1]{\vspace*{0.05in}\noindent\textbf{#1}}
\title{Lightweight and Generalizable Acoustic Scene Representations via Contrastive Fine-Tuning and Distillation}
\name{\parbox{0.8\textwidth}
{\centering Kuang Yuan$^{1,2}$\sthanks{This work was done during an internship at Meta Reality Labs.}, 
Yang Gao$^{1}$, 
Xilin Li$^{1}$, 
Xinhao Mei$^{1}$, \\ 
Syavosh Zadissa$^{1}$, 
Tarun Pruthi$^{1}$, 
Saeed Bagheri Sereshki$^{1}$}}

\address{$^{1}$Meta Reality Labs, Redmond, WA, USA\;
$^{2}$Carnegie Mellon University, Pittsburgh, PA, USA
}

\begin{document}
%
\maketitle
\begin{abstract}
Acoustic scene classification (ASC) models on edge devices typically operate under fixed class assumptions, lacking the transferability needed for real-world applications that require adaptation to new or refined acoustic categories. We propose \sname, which learns generalizable acoustic scene representations by structuring the embedding space to preserve semantic relationships between scenes, enabling adaptation to unseen categories without retraining. Our approach combines supervised contrastive fine-tuning of pre-trained models with contrastive representation distillation to transfer this structured knowledge to compact student models. Our evaluation shows that \sname\ demonstrates improved few-shot adaptation to unseen categories while maintaining strong closed-set performance.
\end{abstract}


%

\section{Introduction}
\label{sec:intro}
Acoustic scene classification (ASC) has attracted significant research attention as a crucial capability for context-aware AI systems on edge devices~\cite{dcase2024,asc_survey2023}. Recent advances have achieved promising results in developing low-resource, on-device ASC systems that balance accuracy with computational efficiency~\cite{dcase2025}. However, these approaches universally operate under the assumption of a fixed set of acoustic scene classes determined at training time, fundamentally lacking the transferability needed for real-world deployment.


The importance of transferability becomes evident in practical applications where acoustic scene understanding must adapt to new or refined categories. Consider a hearing aid that initially identifies broad categories like ``dining venue'' but later needs to differentiate between a quiet restaurant and a bustling cafeteria to optimize noise suppression settings. Similarly, an indoor navigation assistant may initially be trained only on ``bus'' and ``metro'' due to limited dataset diversity, but later need to recognize ``tram'' environments to provide accurate transit guidance. Such scenarios demand ASC models capable of adapting to new acoustic scene categories without complete retraining.

While prior efforts have extensively studied critical challenges in ASC including device mismatch~\cite{device_mismatch2022}, data efficiency~\cite{data_efficient2024}, and model compression, the fundamental challenge of \textit{classification transferability} remains largely unexplored. Traditional supervised training methods employ cross-entropy loss to create highly discriminative features for predefined classes, resulting in embeddings that excel at closed-set classification but fail to generalize to unseen categories~\cite{semantic2024}.

Instead of optimizing for specific class predictions, our approach focuses on learning generalizable acoustic scene representations by structuring the embedding space to preserve semantic relationships between scenes. We achieve this through supervised contrastive learning, which pulls together embeddings of the same acoustic scene while pushing apart different scenes, creating a geometrically structured feature space that naturally supports transferability~\cite{supcon2020}. This contrasts with traditional cross-entropy training that optimizes decision boundaries for fixed classes, often resulting in representations that lack generalization capability.

However, deploying such models on edge devices presents a critical challenge: standard knowledge distillation methods focus on transferring final predictions to compact student models, potentially losing the carefully structured embedding space that enables generalization~\cite{kd_asc2020,teacher_kd2025}. To address this, we employ Contrastive Representation Distillation (CRD)~\cite{crd2020}, which preserves the relational structure between embeddings during model distillation. 


\begin{figure*}
    \centering
    \includegraphics[width=0.9\linewidth]{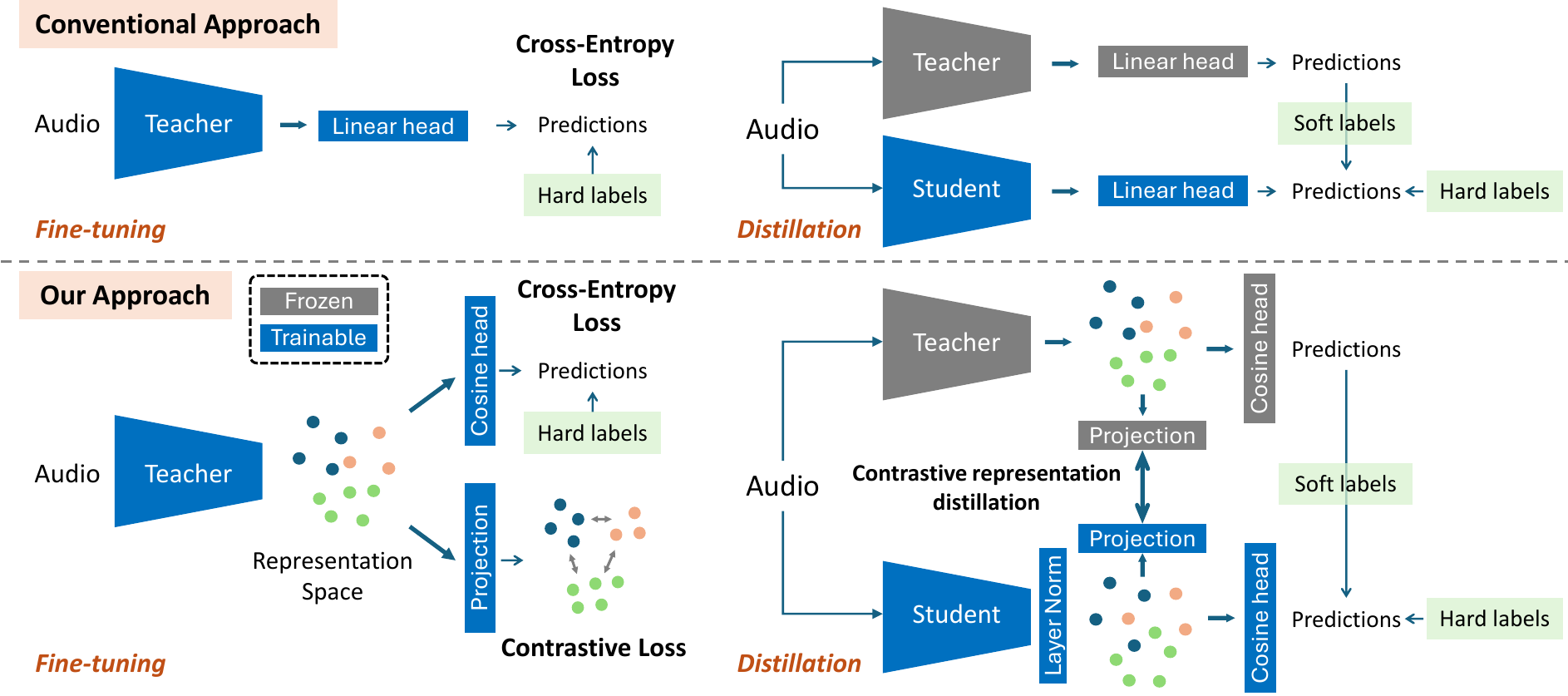}
    \caption{Overview of the proposed 2-stage training framework of \sname\ in contrast to the conventional approach.}
    \label{fig:overview}
\end{figure*}

In this paper, we propose \sname, a two-stage framework that adapts supervised contrastive learning and CRD for lightweight and generalizable acoustic scene representations (see Fig.~\ref{fig:overview}). First, a pre-trained BEATs model~\cite{beats2022} is fine-tuned using a mixup-aware supervised contrastive loss with cosine similarity heads to produce transferable teacher representations that maintain semantic structure. Second, we apply CRD with carefully designed projection layers and layer normalization to transfer the relational structure from the teacher's representations to the embedding space of a compact CP-Mobile model~\cite{cpmobile2023}. We perform training and distillation using the TAU22 dataset~\cite{tau22}. Our evaluation demonstrates that \sname\ achieves strong closed-set performance on TAU22 and superior open-set generalization through few-shot adaptation to other datasets~\cite{tut17,icme2024} with unseen acoustic scenes.

\section{Contrastive Fine-Tuning}
\label{sec:contrastive_finetuning}

Traditional ASC fine-tuning relies on cross-entropy loss, which optimizes decision boundaries for fixed class sets but produces embeddings that struggle to generalize to unseen acoustic categories~\cite{dcase2024, teacher_kd2025, cpmobile2023}. As the first stage of our two-stage framework, we supplement cross-entropy training with supervised contrastive learning to structure the embedding space and preserve semantic relationships between acoustic scenes. This section details the contrastive fine-tuning techniques and training methodology for the teacher model.



\sssec{Model Architecture.}
We employ the pre-trained BEATs model~\cite{beats2022} as our teacher backbone, which learns bidirectional encoder representations from audio transformers through iterative acoustic tokenization. Unlike prior ASC works~\cite{cpmobile2023, teacher_kd2025} that utilize PaSST as the teacher model designed for audio sampled at 32~kHz, BEATs operates on 16~kHz audio, making it more suitable for knowledge transfer to compact edge models that typically process audio at this sampling rate. The BEATs encoder produces 768-dimensional embeddings that capture rich semantic audio representations.

Following established contrastive learning frameworks~\cite{supcon2020}, we augment the BEATs backbone with a two-layer MLP projection head for supervised contrastive learning. The projection head transforms the 768-dimensional BEATs embeddings through two fully connected layers to 128-dimensional projections for contrastive loss computation.

\sssec{Cosine Classification Head.}
In our proposed method, we replace the standard linear classification head with a cosine similarity-based classifier to enhance embedding transferability. The cosine classifier computes logit for each class as $\text{logits}_c = \gamma \, \text{COS}(\mathbf{x}, \mathbf{w}_c)$, where $\text{COS}(\mathbf{x}, \mathbf{w}_c)$ is the cosine similarity between the the input embedding $\mathbf{x}$ and the trainable weight vector $\mathbf{w}_c$ for class $c$, and $\gamma$ is the scale parameter. 
This formulation bounds the logits and reduces sensitivity to embedding magnitudes, promoting more generalizable feature representations~\cite{cosine_norm2017}. We empirically set $\gamma = 56$ based on the validation performance.

\sssec{Mixup-Aware Supervised Contrastive Loss.}
Traditional supervised contrastive learning~\cite{supcon2020} operates on discrete class labels, limiting its compatibility with mixup augmentation~\cite{mixup} which generates interpolated labels through linear combination. To address this limitation, we propose a new mixup-aware supervised contrastive loss that leverages the semantic relationships encoded in mixup's interpolated labels.

For a batch of $B$ samples, our mixup-aware supervised contrastive loss is formulated as:
\begin{equation}
\mathcal{L}_{\text{Soft-SupCon}} = -\frac{1}{B}\sum_{i=1}^B \log \frac{\sum_{k \neq i} w_{ik} \exp(s(i,k))}{\sum_{m \neq i} \exp(s(i,m))} \,,
\label{eq:soft_supcon}
\end{equation}
where $w_{ik}$ represents the similarity weight between samples $i$ and $k$, defined as the dot product of their respective mixup label vectors. The similarity function $s(i,k)$ is the temperature-scaled cosine similarity $s(i,k) = \text{CSF}(\mathbf{z}_i, \mathbf{z}_k) / \tau$, where $\mathbf{z}_i$ and $\mathbf{z}_k$ are the $\ell_2$-normalized projection embeddings, and $\tau$ is the temperature parameter. This formulation enables the contrastive loss to adapt to the continuous nature of mixup labels, pulling together embeddings with higher label similarity while maintaining the contrastive structure.

\sssec{Data Augmentation.}
Our data augmentation pipeline combines time and frequency domain augmentations to improve model robustness. We implement Freq-MixStyle (FMS)~\cite{freqmixstyle}, mixup~\cite{mixup}, frequency masking, and time rolling. Freq-MixStyle applies domain randomization in the frequency dimension by mixing feature statistics across different recordings, simulating new acoustic domains during training. The augmentation parameters used in our training are: FMS probability $p_{\text{FMS}} = 0.4$ with mixing strength $\alpha_{\text{FMS}} = 0.4$, mixup probability of $0.3$ with $\alpha_{\text{mixup}} = 2.0$, time rolling up to $0.1$ seconds, and frequency masking up to $48$ mel bins.

\sssec{Training Procedure.}
Training proceeds in two phases on the TAU22 dataset. First, we freeze the BEATs backbone and independently initialize the cosine classification head and MLP projection head for $50$ epochs using AdamW optimizer ($lr = 0.008$, weight decay $= 10^{-4}$, batch size $= 2048$). Empirically, we found that applying data augmentation only to the projection head while using clean samples for the classification head yields lower validation loss.

Subsequently, we perform joint end-to-end fine-tuning for $30$ epochs with a cosine annealing scheduler with 2 epochs warmup (peak learning rate is $10^{-4}$). The overall loss combines cross-entropy and supervised contrastive objectives:
\begin{equation}
\mathcal{L}_{\text{Fine-tuning}} = \lambda \mathcal{L}_{\text{CE}} + (1-\lambda) \mathcal{L}_{\text{Soft-SupCon}}\,,
\label{eq:total_loss}
\end{equation}
where $\lambda = 0.25$ balances the two objectives, and the temperature parameter $\tau = 0.2$.
\section{Contrastive Representation Distillation}
\label{sec:contrastive_distillation}

Traditional knowledge distillation transfers final predictions, but fails to preserve the structural properties of embeddings that enable generalization. In the section, we detail the second stage of \sname\ about applying Contrastive Representation Distillation (CRD)~\cite{crd2020} to transfer the relational structure from teacher embeddings to compact student models.

\sssec{Student Model Architecture.}
We adopt CP-Mobile~\cite{cpmobile2023} as the student backbone, a state-of-the-art compact model for ASC. In its original form, CP-Mobile concludes with a classification block: Conv2D (mapping the feature dimension to the number of classes) $\rightarrow$ BatchNorm2D $\rightarrow$ AvgPooling2D. To enable explicit embedding extraction, we modify this architecture to: AvgPooling2D $\rightarrow$ LayerNorm1D $\rightarrow$ cosine classification head, where the 1D embeddings are taken after the LayerNorm stage. 

We employ a cosine similarity–based classification head, consistent with the teacher model, to improve transferability. Furthermore, we replace BatchNorm with LayerNorm, which normalizes each sample independently. This design yields more stable representations across varying data distributions, thereby further enhancing transferability~\cite{xu2019understanding}.

To adapt CP-Mobile from its original 32~kHz sampling rate to the 16~kHz rate used in BEATs, we adjust the front-end parameters by halving the window length, hop length, and FFT size. This preserves the temporal resolution, yielding a 96~ms window and 16~ms hop length.

\sssec{CRD Projection Heads.}
The original CRD framework~\cite{crd2020} employs single linear layers as projection heads for both teacher and student models. In contrast, we adopt more expressive 2-layer MLP projection heads to better capture non-linear relationships in the embeddings. For the teacher, we reuse the frozen projection head trained during the contrastive fine-tuning stage. For the student, we initialize a 2-layer MLP that maps its embeddings into the same 128-dimensional space as the teacher projections. Compared to linear mappings, MLP-based projections provide more flexible and effective representation alignment~\cite{chen2020simple}.

\sssec{CRD Loss.} The CRD loss preserves pairwise sample relationships by maximizing the lower bound of mutual information between teacher and student representations:
\begin{equation}
\mathcal{L}_{\text{CRD}} = -\mathbb{E} \left[ \log \frac{\exp(q^T \cdot k^S / \tau)}{\exp(q^T \cdot k^S / \tau) + \sum{k^S_-} \exp(q^T \cdot k^S_- / \tau)} \right]
\end{equation}
where $q^T = h_{\text{proj}}^T(e_t)$ is the teacher's projected embedding, $k^S = h_{\text{proj}}^S(e_s)$ is the student's projected embedding for the positive pair, $k^S_-$ represents negative student samples from different inputs, and $\tau$ is the temperature parameter.






\sssec{Data Augmentation.}
In addition to the augmentations described in Section~\ref{sec:contrastive_finetuning}, we incorporate two further techniques for CRD: (i) device impulse response (DIR) augmentation~\cite{morocutti2023device, yuan2025sonicsieve}, applied with probability 0.6, and (ii) frequency shift augmentation, which randomly shifts the STFT’s maximum frequency range within $\pm$1000 Hz to simulate variations in frequency response. We omit mixup, as CRD requires well-defined positive and negative pairs.

\sssec{Training Procedure.}
We optimize the student using a combined loss function during the distillation stage:
\begin{equation}
\mathcal{L}_{\text{distillation}} = \alpha \mathcal{L}_{\text{CE}} + (1-\alpha) \mathcal{L}_{\text{KD}} + \beta \mathcal{L}_{\text{CRD}}
\label{eq:distillation_loss}
\end{equation}
where $ \mathcal{L}_{\text{KD}}$ is the standard knowledge distilation loss based on KL-divergence. We set $\alpha = 0.02$, $\beta = 0.1$, knowledge distillation temperature $\tau_{\text{KD}} = 2.0$, and the CRD temperature $\tau_{\text{CRD}} = 0.07$. Following the hyperparameters set in CP-Mobile, we train five model variants (6K-126K parameters) using AdamW optimizer with peak learning rates $\{0.04, 0.04, 0.03, 0.02, 0.01\}$, and cosine annealing scheduling over 75 epochs with 7 warmup epochs.
\section{Results}
\label{sec:results}

\begin{table}[t]
\centering
\caption{Accuracy and mAP of teacher models (FT: Fine-tuning; CE-only: Cross-Entropy loss only)}
\label{tab:teacher}
\small 
\setlength{\tabcolsep}{3.5pt} 
\begin{tabular}{c c c c c}
\toprule
\multirow{3}{*}{\textbf{Method}} & \multirow{2}{*}{\textbf{TAU22}} & \multicolumn{3}{c}{\textbf{TUT17 (Open-set)}} \\
\cmidrule(lr){3-5} 
& \textbf{(Close-set)} & \textbf{5-shot} & \textbf{20-shot} & \textbf{mAP}\\
\midrule
BEATs (Frozen) & 55.8 & 55.9 & 67.6 & 0.48\\
FT (CE-only) & \textbf{62.5} & 60.1 & 70.4 & 0.54\\
\midrule
Contrastive FT & \textbf{62.5} & \textbf{62.3} & \textbf{72.4} & \textbf{0.58}\\
Contrastive FT (linear head) & \textbf{62.6} & 61.7 & 72.2 & 0.57\\
\bottomrule
\end{tabular}
\end{table}

\begin{table}[htbp]
\centering
\vspace{-3mm}
\caption{Accuracy of CP-Mobile~(126K) student model and ablation study (C-FT: Contrastive Fine-tuning)}
\label{tab:student}
\small
\setlength{\tabcolsep}{3pt}
\begin{tabular}{c c c c c c c}
\toprule
\multirow{2}{*}{\textbf{Teacher}} & \multirow{2}{*}{\textbf{KD}} & \multirow{2}{*}{\textbf{TAU22}} & \multicolumn{2}{c}{\textbf{TUT17}} & \multicolumn{2}{c}{\textbf{ICME24}} \\
\cmidrule(lr){4-5} \cmidrule(lr){6-7}
& \textbf{method}& \textbf{(Close-set)}& \textbf{5-shot} & \textbf{20-shot} & \textbf{5-shot} & \textbf{20-shot} \\
\midrule
\multicolumn{2}{c}{\textit{No KD}} & 57.4 & 50.7 & 61.2 & 58.2 & 72.2 \\
FT & KD & 59.3 & 53.0 & 62.9 & 62.6 & 75.4 \\
FT & CRD & 60.0 & 55.1 & 65.8 & 61.7 & 76.0 \\
C-FT & KD & 59.9 & 56.1 & 64.5 & 63.5 & 76.1 \\
\bottomrule
\textbf{C-FT} & \textbf{CRD} & \textbf{60.6} & \textbf{56.3} & \textbf{66.5} & \textbf{64.5} & \textbf{77.7} \\
\multicolumn{2}{r}{\textit{- LayerNorm}} & 60.4 & 56.4 & 65.9 & 63.8 & 76.5 \\ 
\multicolumn{2}{r}{\textit{+ BatchNorm}} & 60.0 & 54.9 & 65.8 & 62.2 & 76.3 \\
\bottomrule
\end{tabular}
\end{table}
\sssec{Evaluation Protocol.}
We perform all fine-tuning and distillation on TAU22~\cite{tau22}, and evaluate closed-set performance on its validation split. To further assess open-set transferability, we evaluate the trained encoder on TUT17~\cite{tut17} and ICME24~\cite{icme2024}. Both of the datasets include multiple categories that are unseen during training.  For each dataset, we run a $K$-shot evaluation: sample $K$ training examples per class, extract embeddings using our model, fit a logistic-regression classifier on these $K$-shot embeddings, and evaluate on the full held-out test set. We repeat this sampling 300 times and report mean accuracy for $K\in\{5,20\}$. We note that the pre-LayerNorm embedding is used for few-shot classification, which empirically yields better generalization.

\sssec{Teacher Model Performance.}
Table~\ref{tab:teacher} demonstrates that contrastive fine-tuning significantly improves open-set generalization while maintaining competitive closed-set performance. Compared to standard cross-entropy fine-tuning, contrastive fine-tuning achieves similar TAU22 accuracy (62.5\%) but substantially better few-shot adaptation on TUT17. The mean Average Precision (mAP) metric, which evaluates embedding quality by measuring the overall retrieval performance, further confirms the superior representatios learned through contrastive fine-tuning (0.58 vs 0.54). Notably, the cosine classification head provides marginal improvements of the transferability over the linear head.

\sssec{Knowledge Transfer to Student Models.}
Table~\ref{tab:student} shows that CRD successfully transfers the structured knowledge from contrastive fine-tuned teachers to compact CP-Mobile (126K version). The combination of contrastive fine-tuning with CRD achieves the best performance across all metrics, improving 5-shot accuracy on TUT17 from 53.0\% (FT + KD) to 56.3\% and on ICME24 from 62.6\% to 64.5\%. The architectural ablation study further reveals that our design with LayerNorm consistently outperforms both no normalization and BatchNorm variants, confirming our hypothesis that sample-independent normalization enhances transferability.

\begin{figure}[t]
    \centering
    \includegraphics[width=1\linewidth]{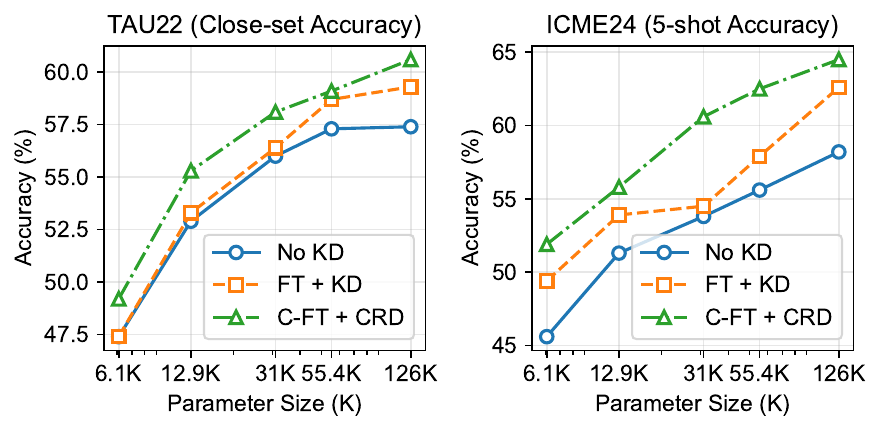}
    \vspace{-8mm}
    \caption{Close-set and open-set accuracies across model sizes}
    \vspace{-2mm}
    \label{fig:model_size}
\end{figure}

\begin{table}[h]
\vspace{-2mm}
\centering
\caption{5-shot accuracy on seen and unseen classes}
\label{tab:classes}
\small
\setlength{\tabcolsep}{6pt}
\begin{tabular}{c c c c c}
\toprule
\multirow{2}{*}{\textbf{Method}} & \multicolumn{2}{c}{\textbf{TUT17}} & \multicolumn{2}{c}{\textbf{ICME24}} \\
\cmidrule(lr){2-3} \cmidrule(lr){4-5}
& \textbf{Seen} & \textbf{Unseen} & \textbf{Seen} & \textbf{Unseen} \\
\midrule
No KD & 44.1 & 53.2 & 59.6 & 57.7 \\
FT + KD & \textbf{48.7} & 54.8 & 64.3 & 61.2 \\
\textbf{C-FT + CRD} & 47.9 & \textbf{59.6} & \textbf{65.8} & \textbf{64.3} \\
\bottomrule
\end{tabular}
\vspace{-3mm}
\end{table}
\sssec{Transferability Analysis.}
Table~\ref{tab:classes} provides insight into the transferability by decomposing the performance on seen and unseen classes. Compared to standard pipeline (FT+KD), our approach (C-FT+CRD) achieves comparable performance on seen classes, while demonstrates superior generalization to unseen categories (3/10 and 11/15 classes are unseen in TUT17 and ICME24, respectively). This highlights the effectiveness of \sname\ for new category adaptation.

\sssec{Scalability Across Model Sizes.}
Figure~\ref{fig:model_size} demonstrates consistent performance improvements across the full range of CP-Mobile variants (6K-126K parameters). Our approach achieves more significant and consistent performance improvements compared to the conventional approach (FT + KD). On closed-set TAU22 evaluation, improvements range from 1.8\% to 3.2\%, while open-set ICME24 gains are even more pronounced, reaching up to 6.3\% improvement. These results confirm that the benefits of structured representation learning scale effectively across different sizes of models.


\section{Conclusions and Future Works}
We present \sname, a two-stage framework that learns lightweight and generalizable acoustic scene representations through contrastive fine-tuning and distillation. \sname\ enables effective few-shot adaptation to unseen categories while maintaining competitive closed-set performance. Future work could integrate teacher ensembling with our approach to further improve representation generalizability.



\pagebreak

\bibliographystyle{IEEEbib}
\bibliography{refs}

\end{document}